\documentclass[a4paper,11pt]{article}
\pdfoutput=1 

\usepackage{jcappub} 

\usepackage[T1]{fontenc} 

\title{\boldmath Primordial reheating in $f(R)$ cosmology by spontaneous decay of scalarons}

\author[a]{Arun Mathew}
\author[b,1]{and Malay K. Nandy\note{Corresponding Author.}}

\affiliation[a]{Dublin Institute for Advanced Studies, Astronomy \& Astrophysics Section, DIAS Dunsink Observatory, Dublin D15 XR2R, Ireland}
\affiliation[b]{Department of Physics, Indian Institute of Technology Guwahati, Guwahati 781039, India}

\emailAdd{arun@cp.dias.ie}
\emailAdd{mknandy@iitg.ac.in}

\abstract{We employ a viable $f(R)$ gravity model capable of giving an inflationary phase in order to study the subsequent reheating phase due to particle creation at the expense of energy in the scalaron field. Since quantum mechanics is expected to play a dominant role in particle creation, we formulate a plausible scenario of reheating obeying Heisenberg's uncertainty principle that imposes constraints on the particles created in the configuration space. We show that, so long as the energy available in the scalaron field is sufficient to populate the entire configuration space, the energy density of the particles grows, attaining a maximum value giving an efficient reheating. Beyond this maximum, the available energy becomes insufficient to populate the entire configuration space leading to a declining energy density.  

We further find that there is a negligible growth of energy density in the inflationary phase that lasts for $\sim 10^7 \, t_{\rm P}$, although particles are constantly created in this phase. The subsequent reheating phase spans for $\sim10^{11} \, t_{\rm P}$ and it begins with a well-defined preheating stage lasting for $\sim 10^{5} \, t_{\rm P}$, making a cross-over to a thermilization regime. The temperature at the beginning of the thermilization is found to be $T_{\rm th}\sim 10^{12}$ GeV, whereas the reheating temperature is estimated as $T_{r}\sim10^{13}$ GeV. Importantly, these estimates follow from a single parameter, the  scalaron mass, $M\sim10^{-5} \, M_{\rm P}$.
}

\begin{document}
\maketitle
\flushbottom

\section{Introduction}

The standard Friedmann model of cosmology \cite{Weinberg2008,Peebles1993} has explained several observed features of our Universe such as primordial nucleosynthesis \cite{Fields1996}, abundances of the light elements \cite{Schramm1993}, the cosmic microwave background (CMB) radiation \cite{Peebles1991,Mather1994}, the Hubble expansion \cite{Tammann1992}, apart from the fact that it has the well-known problem of the Big-Bang singularity. Moreover, after the observation of CMB radiation \cite{Penzias1965},  the standard Friedmann model failed in several aspects giving rise to the monopole problem, the horizon problem, the flatness problem and the problem of large-scale structure formation \cite{Weinberg2008,Kolb1990}.  To solve these problems, it was suggested by Guth \cite{Guth1981}, Linde \cite{LInde1982a,LInde1982b,LInde1982c}, Albrecht and Steinhardt \cite{Albrecht1982} and others \cite{Starobinsky1980, Starobinsky1979} that the universe underwent a fast exponential expansion, dubbed inflation, prior to the radiation dominated phase. A lot of efforts have been expended in understanding the inflationary era \cite{Linde1984}. While Guth connected it to a first-order phase transition in a Grand Unified Theory,  Linde and others \cite{Linde1983,Linde1984,Linde1985} proposed slow-roll models with one or more inflaton fields driving the inflationary phase. Moreover, these theories can explain the subsequent radiation era via a reheating phase where the inflaton field decays to generate Standard Model particles \cite{Dolgov1989,Traschen1990, Kofman1994,Shtanov1995,Kofman1997,Khlebnikov1996}. It is however not clear at the outset the intrinsic origin of the postulated inflaton field. Although initially the inflaton field was understood to be the Higgs field of the Standard Model, it was later realised that observational data did not fit this assumption \cite{Bezrukov2008}.

Interestingly, Satrobynsky \cite{Starobinsky1980} had proposed a higher order theory of gravity that addressed the problem of Big-Bang singularity. He considered the Einstein field equation $G_{\mu\nu}=\kappa \langle T_{\mu\nu}^{\textsc{qm}} \rangle $ with the right-hand side given by the vacuum expectation value due to quantum matter fields (having different spins) in the background of the classical gravitational field, with the assumption of  isotropy and homogeneity in the absence of radiation. The vacuum expectation value $\langle T_{\mu\nu}^{\textsc{qm}} \rangle $ is determined by Riemann geometric quantities in the one-loop approximation \cite{Davies1977,Bunch1977,Birrell1982}. The ensuing solution was found to be of non-singular de-Sitter type that analytically continues to the region $t<0$. Thus an inflationary scenario, in the form of a de-Sitter phase,  follows without the need for an inflaton field in the theory. 

In addition, it is important to understand the origin of the radiation dominated era following the inflationary phase. In fact, there have been a few studies on particle creation due to varying gravitational field \cite{Parker1968,Parker1969,Parker1971,Zeldovich1971,Zeldovich1977,Brout1980}. Importantly, it was found that the particle production rate depends on the invariants of the Riemann curvature tensor for the creation of massless scalar particles in a weak Bianchi type-one metric \cite{Zeldovich1977}. For instance, the production rate of massless scalar particles in a weakly anisotropic metric goes like  $C_{\alpha \beta \gamma \sigma} C^{\alpha \beta \gamma \sigma}$, where $C_{\alpha \beta \gamma \sigma}$ is the Weyl tensor. An analogous rate for the production of photons and neutrinos also holds true \cite{Zeldovich1977}.  It was further shown that the graviton production rate in an isotropic metric is proportional to the square of the Ricci scalar \cite{Zeldovich1977}. Moreover, in the Starobynsky scenario, it is demonstrated that particles are created via decay of scalarons during the rapid oscillatory phase following the inflationary regime \cite{Starobinsky1984}, giving rise to a thermalized radiation dominated era.

Ford \cite{Ford1987} considered creation of particles due to a transition from a de-Sitter phase to a Robertson-Walker universe. He suggested that gravitational particle creation could be a dominant mechanism when the damping of the inflaton field is inefficient in the new inflation models that suffer from the difficulty of inefficient reheating in order to meet two conflicting criteria, namely, a sufficient inflation along with small density fluctuations, requiring a weak coupling between the inflaton and matter fields. 

Mijic et. al. \cite{Mijic1986} considered a gravitational Lagrangian augmented by a quadratic term $\epsilon R^2$. This leads to a rapid oscillatory phase following the inflationary phase that may be viewed as coherent oscillations \cite{Vilenkin1985}, collectively called sclarons, which are responsible for the creation of particles by exciting the matter fields. Considering a scalar matter field, it was found that the amplitude of the wave function follows a Schr\"{o}dinger-like equation with the potential determined by the scale factor and the Ricci scalar \cite{Mijic1986}. An approximate solution for the particle production rate due to transition between initial and later stages turned out to be proportional to $R^2/\sqrt{\epsilon}$ having a $\pi/2$ phase shift in $R$. Identifying the reheating temperature with the energy density created in the early few oscillations led to bounds on the parameter $\epsilon$ by requiring that the reheating temperature must lie between two extremes determined by baryogengesis and GUT phase transition. This led to a wide window for the value of $\epsilon$ which has the possibility of being narrowed down by considering perturbations in the inflationary phase.

Appleby et al. \cite{Appleby2010} considered dark energy $f(R)$ models that suffer from a weak singularity problem with the issue of the scalaron mass overshooting the Planck mass. They considered curing this pathological conditions by adding a term quadratic in the Ricci scalar. Thus the full model could address a primordial inflationary phase along with a late-time de-Sitter phase. However, this combined model was found to give insufficient reheating with a slightly different primordial power spectrum index. The behaviour in the primordial period is different from the original $R^2$ inflationary model. This suggests a profound influence of such modified dark energy models on the dynamics in the primordial phase. Their numerical solutions show abrupt jumps and very few oscillations following the inflationary phase which is a completely different behaviour from the standard oscillatory phase of the pure $R^2$ model. Moreover, this regime indicates inadequate reheating although the growth rate goes like $a \sim t^{1/2}$. This may signal that the universe goes into the present de-Sitter phase after an inflationary phase without an intermediate matter dominant era. Upon considering the back-reaction, the  behaviour was found to be significantly different from those without back-reaction, but the conclusions remained almost the same as before. Whereas their work was carried out in the Jordan frame, Motohashi and Nishizawa \cite{Motohashi2012} analyzed the same problem in the Einstein frame and gave constraints on the parameters involved.

In this paper, we consider the modified gravity model $f(R)=R+\alpha R^2 + \beta R^2 \ln R/\mu^2$ in the Jordan frame. This functional form originates from considering one-loop vacuum fluctuations for a scalar field coupled to an electromagnetic field living in a de-Sitter spacetime \cite{Shore1980}. This functional form nearly corresponds to a Robertson-Walker spacetime (with varying curvature) upto a good approximation for $\alpha\gg\beta$ \cite{Vilenkin1985}. Following Starobinsky \cite{Starobinsky1984}, we implement particle creation via spontaneous decay of the scalaron field. Moreover, we employ the Heisenberg uncertainty principle to estimate the maximum number of particles that can be created in the configuration space of the Universe. We find that the energy available in the scalaron field is sufficient to completely populate the configuration space in the inflationary and reheating regimes, giving rise to an approximately constant rate of energy source in these phases. The reheating phase ends when the energy available in the scalaron field is insufficient to completely populate the configuration space. 

Our present formulation leads to a well-defined preheating stage where the created particles cannot attain thermal equilibrium as their collision rate, determined by their mean free path, lags behind the Hubble rate. When these two rates equalize, transition to a thermilization regime occurs. In subsequent times, the particles attain thermal equilibrium as their collision rate exceeds the Hubble rate. 

It is interesting to note that the scalaron mass $M$ (equivalently, $\alpha$) and the logarithmic prefactor $\beta$ are the only free parameters in our present formulation. Once these parameters are fixed from observational data, the reheating temperature $T_r$ can be predicted. We expect $T_r$ to lie below the energy scale of Grand Unified Theory (GUT) \cite{Pati1973,Georgi1974a,Georgi1974b} since magnetic monopoles are created in the phase transition GUT $\rightarrow$ SU(3) $\times$ SU(2) $\times$ U(1) by spontaneous symmetry breaking around $10^{15}$ GeV \cite{Kibble1980}. We find that this constraint on the energy scale is respected, since the reheating temperature turns out to be $T_r\sim 10^{13}$ GeV upon employing the observational values of the key parameters $\alpha$ and $\beta$.

The remainder of the paper is organized as follows. In Section \ref{f_R_model}, we describe the field equations following from a viable $f(R)$ gravity model capable of  giving an inflationary phase. In Section \ref{Particle_model}, we formulate a model of particle creation based on Heisenberg's uncertainty principle. There, we also formulate the source term responsible for the growth of energy density following from the model of particle creation. We present in Section \ref{Inf_analysis} an analysis of the inflationary phase where we also fix the parameter $\beta$ from the observed CMB anisotropy. The main focus of the work is given in Section \ref{Reh_analysis} where a detailed analysis of the reheating phase is given that follows from the present $f(R)$ gravity model coupled with the model of particle creation. There, we investigate the region of reheating in detail in three different stages, namely, the beginning, the intermediate, and the last stages of the reheating phase. Moreover, in Section \ref{Reh_analysis}, the preheating and thermilization stages are identified by estimating the growth in the collision rate between the particles created in the reheating process. Finally Section \ref{Conclusion} gives a conclusion in relation to the findings in this paper.

\section{A Viable $f(R)$ Gravity Model}\label{f_R_model} 

In this paper,  we shall consider an effective gravitational action in the form of $f(R)$ gravity, given by 
\begin{equation}\label{Action}
S_{\rm grav} = \frac{c^3}{16\pi G} \int d^4 x \ \sqrt{-g} \ f(R), 
\end{equation}
where $R$ is the Ricci scalar. Although the functional form of $f(R)$ can be arbitrary, we shall restrict its form suggested by quantum field theoretical calculations in curved spacetime \cite{Davies1977,Bunch1977,Shore1980,Birrell1982}. Since we are interested in the inflationary and reheating phases, it is important to choose a functional form of $f(R)$ which is relevant, albeit approximate, for these phases of the Universe. 

For a de-Sitter spacetime with constant curvature, where a minimal conformal scalar field is coupled to an electromagnetic field, it was shown by Shore \cite{Shore1980} that the one-loop vacuum fluctuations give rise to an effective potential that may be equivalently written as 
\begin{equation}\label{f_R}
f(R) = R + \alpha R^2 + \beta R^2\ln\frac{R}{\mu^2}, 
\end{equation}
where $\mu$ is the scale of renormalization. Although this form is obtained for a strictly de-Sitter spacetime, it has approximate correspondence to the vacuum fluctuations calculated in a Robertson-Walker metric, namely, that giving rise to the vacuum expectation value of the energy-momentum tensor $\langle T_{\mu\nu}^{\textsc{qm}}\rangle$ due to quantum fluctuations of free, massless, conformally invariant scalar fields, as shown in Refs. \cite{Davies1977,Bunch1977,Birrell1982}. This gives rise to an effective Einstein field equation, $G_{\mu\nu}=\kappa \langle T_{\mu\nu}^{\textsc{qm}}\rangle$, with $\langle T_{\mu\nu}^{\textsc{qm}} \rangle = k_1 \ H_{\mu\nu} ^{(1)} + k_3 \ H_{\mu\nu}^{(3)} $ where $\kappa =8\pi G/c^4$, $k_1$ is an arbitrary constant and $k_3$ is determined by the number of quantum degrees of freedom. In fact, $H_{\mu\nu} ^{(1)}$ can be directly obtained from the $R^2$ term in $f(R)$, whereas, $H_{\mu\nu}^{(3)} $ cannot be obtained by varying the action. Vilenkin \cite{Vilenkin1985} showed that the functional form of $f(R)$ given by (\ref{f_R}) approximately yields the above field equation for $\alpha \gg \beta $ in the case when the curvature of the background spacetime is not constant.  

The field equation for a general $f(R)$ gravity with a matter Lagrangian takes the form 
\begin{equation}\label{General_f_R_FE}
f_R R_{\mu\nu} -\frac{1}{2}fg_{\mu\nu} + (g_{\mu\nu} \nabla^\alpha \nabla_\alpha-\nabla_\mu\nabla_\nu)f_R = \kappa T_{\mu\nu}, 
\end{equation}
where $f_R =\partial f/\partial R$ and $T_{\mu\nu}$ is the energy-momentum tensor of matter given by $T_{\mu\nu} = (\rho+p)u_{\mu}u_{\nu}-pg_{\mu\nu}$, with $\rho$ and $p$ being the proper energy density and proper pressure, respectively. Substituting the functional form of $f(R)$ from equation (\ref{f_R}) in equation (\ref{General_f_R_FE}), we obtain the field equation as
\begin{multline}\small
G_{\mu\nu}  + \beta R R_{\mu\nu} +\left\{\alpha R + \beta R \ln \frac{R}{\mu^2} \right\}\left\{ 2 R_{\mu\nu}-\frac{1}{2}Rg_{\mu\nu}\right\} \\
+ (g_{\mu\nu} \Box  - \nabla_\mu \nabla_\nu) \left\{ 2\alpha R + \beta R + 2\beta R \ln \frac{R}{\mu^2}\right\} 
= \kappa T_{\mu\nu}, 
\end{multline}
where $G_{\mu\nu}=R_{\mu\nu}-\frac{1}{2}g_{\mu\nu}R$ is the Einstein tensor. 

It was shown by Vilenkin \cite{Vilenkin1985} that  a long quasi-de-Sitter (or inflationary) phase can be obtained when the parameters are such that $\alpha\gg\beta$. Employing this approximation, the above field equation reduces to 
\begin{equation}\label{FE}
G_{\mu\nu}  + \beta R R_{\mu\nu} +\alpha R \left( 2 R_{\mu\nu}-\frac{1}{2}Rg_{\mu\nu}\right) 
+ 2\alpha(g_{\mu\nu} \Box - \nabla_\mu \nabla_\nu) R = \kappa T_{\mu\nu},  
\end{equation}
with the corresponding trace equation as 
\begin{equation}\label{Trace_Equation}
6\alpha \Box R  + \beta R^2-R = \kappa T.
\end{equation}

We shall assume a Friedmann-Lema\^{i}tre-Robertson-Walker metric \cite{Landau1982,Weinberg1972} given by
\begin{equation}\small \label{metric}
ds^2 = - dt^2 + a^2(t) \left[ \frac{dr^2}{1-kr^2} + r^2 (d\theta^2 + \sin^2 \theta \ d\varphi^2)\right],
\end{equation}
where $k$ is the curvature parameter. 

The $00$-component of the field equation (\ref{FE}) yields the dynamics of the Hubble rate $H=\frac{\dot{a}}{a}$ as 
\begin{multline}\label{k_include}
3\left\{H^2 + \frac{k}{a^2}\right\}  - 18 \beta\left\{ \dot{H} + 2 H^2 +\frac{k}{a^2} \right\} \left\{\dot{H} + H^2 \right\}
-18\alpha \left\{ \dot{H} + 2 H^2 +\frac{k}{a^2} \right\} \left\{ \dot{H} - \frac{k}{a^2}\right\} \\
+ 36\alpha H \left\{\ddot{H} + 4 H \dot{H} - 2H \frac{k}{a^2} \right\} = \kappa \rho. 
\end{multline}

Since we are interested in a long de-Sitter phase in the initial part of the evolution, this signifies an exponential growth of the scale factor $a(t)$. This implies that the $k$-dependent terms in equation (\ref{k_include}) are insignificant in determining the dynamics. Neglecting these terms, equation (\ref{k_include}) further reduces to 
\begin{equation}\label{H_Equation}
3H^2  - 18 \beta\left(2 H^4 + 3\dot{H} H^2 + \dot{H}^2 \right)\\+ 18\alpha \left( 2\ddot{H} H + 6 \dot{H} H^2 -\dot{H}^2\right)= \kappa \rho,
\end{equation}
which will be referred to as the field equation in the following.

The covariant derivative of the field equation (\ref{FE}) gives $\nabla_\nu T^{\mu\nu}=0$, which in the metric (\ref{metric}), yields 
\begin{equation}\label{}
\dot{\rho} + 3 H (\rho + p) = 0. 
\end{equation}
Further, we shall assume that the matter is described by the equation of state $p=\omega\rho$, reducing the above equation to 
\begin{equation}\label{rho_equation}
\dot{\rho}  + 3H(1 + \omega)\rho = 0. 
\end{equation}
This equation expresses conservation of matter in the absence of any source. Since we start with an action without a non-trivial interaction between the curvature field and matter, the source term is not included in the dynamics.

We shall see that the scalar curvature falls off quickly in the inflationary phase and thereafter it performs a damped oscillation. This oscillation, dubbed scalaron, is coherent in nature \cite{Vilenkin1985}, and it is a consequence of the $\alpha R^2$ term in the Lagrangian. As shown by Birrell and Davies \cite{Birrell1982}, the $\alpha R^2$ term is an effective contribution from the quantum vacuum fluctuations of matter fields coupled to the Ricci scalar. Moreover, as shown by Starobinsky \cite{Starobinsky1984}, the energy density of the scalarons acts like a source  for spontaneous creation of particles. 

In the next Section, we shall model this spontaneous decay of scalarons employing the Heisenberg uncertainty principle.

\section{A Model of Particle Creation}\label{Particle_model}

Assuming the inflation to start near the GUT scale $\sim \ell_{\textsc{gut}}$, the Universe undergoes a quasi-de-Sitter expansion so that the size of universe grows as $\ell(t)=\ell_{\textsc{gut}} \ e^{N(t)}$, where $N(t)$ is the number of e-foldings. The corresponding volume of the universe $V(t)\sim\ell^3(t)$ also undergoes a similar expansion. Eventually, at time $t=t_e$, when $N(t_e)\approx60$, the inflationary phase ends and an oscillatory phase begins, where the Hubble rate $H(t)$ and the scalar curvature $R(t)$ undergo damped oscillations. Starobinsky showed that  particles are created by means of decay of scalarons in this oscillatory phase. The average energy density of the scalaron field $\langle \rho_R\rangle$ was found to be $\langle \rho_R\rangle = \alpha m_{\rm P}^2\langle R^2 \rangle/16\pi$, (see discussions following equation (38) in Ref.~\cite{Starobinsky1984}).

It may be emphasised that the particle creation is essentially a quantum mechanical phenomenon. Although our present formulation is classical in nature, the quantum nature of particle production must be taken into account, albeit in an approximate fashion. To formulate the particle production, we employ the Heisenberg uncertainty principle, $\Delta x \ \Delta p_x \geq \hbar/2$. In a pair production of ultra-relativistic particles each having energy $\varepsilon$ (on an average), the corresponding particle momentum uncertainty $\Delta p_x \sim  \varepsilon/\sqrt{3}c$ gives a minimum position uncertainty $\Delta x \sim \hbar/2\Delta p_x \sim \sqrt{3}\hbar c/2\varepsilon$. Such particle production is expected to happen throughout the volume $V(t)$ of the Universe. For simplicity, we work with $\varepsilon$ which is the average energy per particle created, so that we may partition the volume $V(t)$ into $\mathcal{N}(t) = V(t)/(\Delta x)^3$ cells. If sufficient energy is available in the scalaron field,  all $\mathcal{N}(t)$ cells can be filled up by creation of $\mathcal{N}(t)$ particles: this scalaron decay process happens within a time-span of $\Gamma^{-1}$, where $\Gamma$ is the decay rate of scalarons. It is important to note that no more than $\mathcal{N}(t)$ particles can be created within the time scale $\Gamma^{-1}$ even if an excessive amount of energy resides in the scalaron field. This restriction is a direct consequence of quantum mechanics via the uncertainty principle. The number density of such created particles is given by $n(t) = \mathcal{N}(t)/V(t) \sim (\Delta x)^{-3}\sim 8\varepsilon^3/3\sqrt{3}\hbar^3 c^3$ with the corresponding energy density $\varepsilon \, n(t)\sim 8\varepsilon^4/3\sqrt{3}\hbar^3 c^3$.  Since particles are created from scalarons with the decay rate $\Gamma$, the production rate $\lambda(t)$ of the energy density, setting $\hbar=c=1$, is given by
\begin{equation}\label{const_condition}
\lambda(t)= \Gamma \, \varepsilon \, n(t) = \frac{8}{3\sqrt{3}}\Gamma \, \varepsilon^4,
\end{equation} 
so long as sufficient energy density $\rho_R$ is available in the scalaron field. The corresponding {\em sufficiency condition} $\rho_R \geq 8\varepsilon^4/3\sqrt{3}$ translates to 
\begin{equation}\label{condition}
 \frac{m_{\rm P}^2}{16\pi}\alpha  R^2\geq \frac{8}{3\sqrt{3}}\varepsilon^4.
\end{equation}
 
In formulating the above scenario, although we employed the average energy $\varepsilon $ per particle, it is expected to give reasonable estimates for macroscopic quantities. Within every decay time $\Gamma^{-1}$, the volume $V(t)$ is populated with particles of different energies $\tilde{\varepsilon}$ with a probability determined by quantum mechanics. In the present formulation, we are interested in the time evolution of the macroscopic energy density $\rho(t)$, which is assumed to be uniform throughout the volume $V(t)$ by the postulate of homogeneity and isotropy of the Universe. Moreover, since the energy density is a macroscopic quantity, its rate of change is slower than the decay rate $\Gamma$. Consequently, it is reasonable to take the average energy $\varepsilon$ of all particles created within the time-span of inverse decay rate $\Gamma^{-1}$.

Since the scalar curvature $R(t)$ decreases in the course of evolution, the equality in sufficiency condition (\ref{condition}) will be reached at some point of time $t_*$. From that time onwards, that is for $t>t_*$, available energy in the scalarons field will be insufficient, and consequently only a fraction $\zeta(t) = \rho_R/(8\varepsilon^4/3\sqrt{3})$ of the $\mathcal{N}(t)$ cells can be filled up
\begin{equation}
\zeta(t) =  \frac{3\sqrt{3}}{128\pi} \frac{\alpha m_{\rm P}^2R^2}{\varepsilon^4},
\end{equation}
so that the particle production rate becomes 
\begin{equation}\label{after_condition}
\lambda(t>t_*) =  \frac{8}{3\sqrt{3}}\Gamma \varepsilon^4  \zeta(t) =   \frac{m_{\rm P}^2}{16\pi}\Gamma \alpha R^2.
\end{equation}
 
In order to account for the particle creation in the macroscopic dynamics, the energy density equation (\ref{rho_equation}) has to be modified by including the particle production rate $\lambda(t)$ as a source term, leading to 
\begin{equation}\label{rho_D_Equation}
\dot{\rho} + 3H (1 + \omega)\rho =  \lambda(t). 
\end{equation}

The corresponding back-reaction is accounted for by modifying the trace equation (\ref{Trace_Equation}) to 
\begin{equation}\small\label{modi_trace_equation}
6\alpha (\ddot{R}+3H\dot{R})  - \beta R^2+R = \kappa \frac{\lambda(t)}{H}-\kappa T, 
\end{equation}
for the case of Friedmann-Lema\^{i}tre-Robertson-Walker metric,  where $T=(3\omega-1)\rho$ is the trace of the energy-momentum tensor.

Due to the above modification in the dynamics, particles will be created both in the inflationary and reheating phases. However, the contribution due to particle production is expected to be insignificant in the inflationary phase due to the quasi-de-Sitter expansion.

\section{Analysis of the Inflationary Phase}\label{Inf_analysis}

In this section, we investigate the region of inflation by imposing the slow-roll approximation in the field equation (\ref{H_Equation}). In this region, the value of the Hubble parameter is expected to be large and is a slowly varying function  given by the slow-roll condition $|\dot{H}|\ll H^2$. Defining a time-scale $\tau$ as $\rho/\tau\sim\lambda$, or $\tau \sim \rho/\lambda $, then $|\dot{H}|\sim H/\tau\sim H \lambda/\rho\ll H^2$ implies $H\rho\gg \lambda$. This implies $\dot{\rho} + 3H (1 + \omega)\rho \approx 0$ in the slow roll regime so that there is no significant growth in the energy density and thus its back-reaction on the background spacetime is negligibly small. Under these considerations, the field equation (\ref{H_Equation}) reduces to 
\begin{equation}\label{H_Diff_Equation_Inf}
\dot{H}   - \frac{1}{3}\frac{\beta}{\alpha} H^2 + \frac{1}{36\alpha} \approx0.
\end{equation}
For a pure de-Sitter phase, the scalar curvature is a constant, say $R_S$, which in this case represents a constant Hubble rate $H_S$. From equation (\ref{H_Diff_Equation_Inf}), we obtain $12 H_S^2=1/\beta$. 
Thus, for a quasi-de-Sitter phase, the second and the third terms are of the same order. Consequently,  the  solution is given by 
\begin{equation}
H(t) = \frac{1}{\sqrt{12\beta}} \tanh \left( \sqrt{12\beta} c_1 - \frac{\sqrt{12\beta}}{36\alpha} t \right),
\end{equation}
where the constant $c_1$ is determined by the initial condition. Rewriting the above solution in terms of  $H_S$, we obtain 
\begin{equation}\label{Pre_Inflation_H}
H(t) =H_S \tanh \left( \frac{c_1}{H_S} - \frac{t}{36\alpha H_S} \right).
\end{equation}
The validity of the above expression is well-founded so far as the slow-roll condition $|\dot{H}|\ll H^2$ is satisfied.

Using equation (\ref{Pre_Inflation_H}), the slow-roll condition gives $1/(6H_S\sqrt{\alpha}) \ll \sinh (c_1/H_S - t/36H_S\alpha )$. Since, the right-hand side is a decreasing function of time, we may guess that at some later time $t_e$, the inequality is no longer valid. This inequality is valid as long as the condition $c_1 \gg t/36\alpha$ is satisfied and one could identify $36 \alpha c_1=t_e$. Hence the condition $|\dot{H}|\ll H^2$ is satisfied for  $t\ll t_e$. Consequently, the end of inflation, given by $|\dot{H}| \approx H^2$, is met when $t\approx t_e$. The expression for Hubble's rate in the region of inflation \cite{Starobinsky1983} is thus obtained as 
\begin{equation}\label{H_Inf}
H(t) =H_S \tanh \left( \frac{t_e-t}{36\alpha H_S} \right).
\end{equation}
From the above equation, one may immediately obtain $t_e$, if the initial value of the Hubble parameter $H_0$ at $t=t_0$ is known. For convenience, we shall set $t_0=0$ in the following. This yields
\begin{equation}\label{t_Inf_E}
t_e = 18 \alpha H_0\ln \left( \frac{H_S+H_0}{H_S-H_0} \right).
\end{equation}
The period of inflation $t_e$ thus depends on the initial value of the Hubble rate $H_0$, and also on the values of the constant parameters $\alpha$ and $H_S=1/\sqrt{12\beta}$. 

Substituting equation (\ref{H_Inf}) in $R = 6(\dot{H}+2H^2)$, we obtain the scalar curvature during inflation as 
\begin{equation}\label{R_Inf}
R = 12H^2 + \frac{1}{6\alpha}\left( \frac{H^2}{H_S^2}-1\right).
\end{equation}

The initial value of the Hubble rate $H_0$ can be fixed from the number of e-foldings during the period of inflation. Number of e-foldings between time $t_0$ and $t$ is given by 
\begin{equation}\label{General_N}
N(t)= \int_{t_0}^{t} H(t) \, dt. 
\end{equation}
Consequently, $N_e$ is obtained by substituting equation~(\ref{H_Inf}) in (\ref{General_N}) so that 
\begin{equation}
N_e = H_S \int_{0}^{t_e} \tanh \left( \frac{t_e-t}{36H_S\alpha} \right) dt,
\end{equation}
which gives
\begin{equation}
N_e =  36\alpha H_S^2 \ln \left\{   \cosh \left(\frac{t_e}{36H_S\alpha}\right) \right\}.
\end{equation}

Using equation (\ref{t_Inf_E}), one may obtain a constraint connecting all parameters of the theory with the initial Hubble rate $H_0$,
\begin{equation}\label{H_0}
H_0 =  H_S \sqrt{1- e^{-\frac{2\beta}{3\alpha}N_e }}.
\end{equation}

As stated earlier, the end of inflation is marked by the condition $|\dot{H}|\sim H^2$ so that 
\begin{equation}\label{H_e}
H_e \sim \frac{1}{6\sqrt{\alpha}},
\end{equation}
and the corresponding scalar curvature is obtained by substituting (\ref{H_e}) in (\ref{R_Inf}), giving
\begin{equation}\label{}
R_e \sim \frac{1}{6\alpha}.
\end{equation}
Thus, during the inflationary phase, the scalar curvature decreases from $R_0$ to $R_e$ and the Hubble rate decreases from $H_0$ to $H_e$. 

The scalaron energy density at the end of inflation, given by $\rho_{R} = \alpha m_{\rm P}^2 R^2/16\pi$, becomes $m_{\rm P}^2/576\pi\alpha$. Since $\alpha$ is related to the scalaron mass $M$ as $\alpha = 1/6M^2$, the scalaron energy density becomes $\rho_R =m_{\rm P}^2 M^2/96\pi$, which is much larger than the energy density of the  particles, $8\varepsilon^4/3\sqrt{3} = M^4/6\sqrt{3}$, since $M=2\varepsilon$ for pair production\footnote{We assume pair production whereby a scalaron particle decays into two bosons. This is similar to the inflaton decaying into two bosons, $\phi\rightarrow \chi\chi$,  which dominates over the decay into two fermions, $\phi\rightarrow \bar{\psi}\psi$ \cite{Linde_Book_1990}.} and $M\ll m_{\rm P}$. This indicates that, during the course of inflation,  the decay rate $\lambda$ would take the form $\lambda=8\Gamma \varepsilon^4/3\sqrt{3}={\rm const}$, as the sufficiency condition (\ref{condition}) holds. Thus from (\ref{rho_D_Equation}), we can immediately write the solution as
\begin{equation}\small\label{rho_Equation}
\rho(t) =  \bigg\{ \rho_0 + u(t) \bigg\}  \  \exp{\left\{-3(1+\omega)\int_0^t H(t') dt'\right\}},
\end{equation}
where  $\rho_0$ is the initial seed density at $t=t_0=0$ and $u(t)$ is given by the integral 
\begin{equation}\small\label{J}
u(t) = \int_0^{t} \lambda(t') \ \exp{ \left\{ 3(1+\omega)\int_0^{t'} H(t'') \ dt''\right\} } \  dt'  .
\end{equation}
The evolution of density during the inflationary phase can be obtained by substituting equation (\ref{H_Inf}) in (\ref{rho_Equation}), so that the integral $u(t)$ can be expressed as
\begin{equation}\small\label{}
u(t) = \frac{8}{3\sqrt{3}}\Gamma \varepsilon^4 \int_{0}^{t}  dt' \exp{\left\{ 3(1+\omega)H_S \int_{0}^{t'} \tanh \left( \frac{t_e -t''}{36\alpha H_S} \right) dt''\right\}},  
\end{equation}
giving
\begin{equation}
u(t) = \frac{8}{3\sqrt{3}}\Gamma \varepsilon^4  \left[\frac{H_S}{H_S-H}\right]^\gamma \int_{0}^{t}  \left\{ 1- \tanh \left( \frac{t_e-t'}{36\alpha H_S}\right)  \right\}^\gamma  dt', 
\end{equation}
where $\gamma = \frac{9\alpha}{\beta}$.

Substituting this expression in equation (\ref{rho_Equation}), we obtain 
\begin{equation}\label{}
\rho (t) = \rho_0  \left( \frac{H_S-H_0}{H_S-H(t)}\right)^\gamma+  \left( \frac{H_S}{H_S-H(t)}\right)^\gamma \left(\frac{8}{3\sqrt{3}}\right)\Gamma \varepsilon^4  \int_{0}^{t}  \left\{ 1- \tanh \left( \frac{t_e-t'}{36\alpha H_S}\right)  \right\}^\gamma  dt'.   
\end{equation}
By the end of inflation, the Hubble rate  $H_e\ll H_S$. Consequently, the first term becomes insignificant by the end of inflation even when the initial seed density $\rho_0\neq0$.  Thus any energy distribution (or seed) present initially is diluted infinitely in the inflationary era. On the other hand, the second term, coming from the source $\lambda$, contributes to the energy density during the inflationary phase. Thus the expression for energy density at the end of inflation $t=t_e$ can be approximated as 
\begin{equation}\small\label{}
\rho(t_e) =  \frac{8}{3\sqrt{3}}\Gamma \varepsilon^4 \left( \frac{2H_S}{H_S-H(t_e)}\right)^\gamma \int_{0}^{t_e}  \left( \exp{\left[ \frac{t_e-t'}{18\alpha H_S}\right]} +1 \right)^{-\gamma} dt',   
\end{equation}
so that 
\begin{equation}\small
\rho_e =  \frac{8}{\sqrt{3}}\beta H_S \Gamma \varepsilon^4 \left( \frac{2H_S}{H_S-H_e}\right)^\gamma  e^{- \frac{(t_e-t)}{3 \beta H_S} } \\ 
\times \ _2 F_1 \left( \gamma, \gamma; \gamma+1; - e^{- \frac{(t_e-t)}{18 \alpha H_S } } \right) \bigg|_0^{t_e},
\end{equation}
where $_2 F_1$ is the hypergeometric function. Since $H_e\ll H_S$, and setting $N_e = 60$, we obtain an approximate value for $\rho_e$ as
\begin{equation}\label{rho_e}
\rho_e \approx  \frac{4\sqrt{\beta}}{3} \Gamma \varepsilon^4 = \frac{2}{3\sqrt{3}}\left(\frac{\Gamma}{H_S}\right) \varepsilon^4,
\end{equation}

Since the particle production rate $\Gamma$ is expected to be much smaller than the excessively large Hubble rate $H_S$ in the de-Sitter phase, $\rho_e$ is negligibly small.

\subsection{Determination of the Parameter $\beta$ from CMB Anisotropy}

In contrast with Einstein's gravity, $f(R)$ gravity has additional slow roll parameters \cite{Felice2010}. In the Jordan frame, the non-vanishing slow-roll parameters are given by 
\begin{equation}
\varepsilon_1 = - \frac{\dot{H}}{H^2},   \hspace{0.4cm} \varepsilon_3 = \frac{\dot{f_R}}{2Hf_R}, \hspace{0.4cm} \varepsilon_4 = \frac{\ddot{f_R}}{H\dot{f_R}}  
\end{equation}

Employing equation (\ref{H_Inf}), we obtain
\begin{equation}
\varepsilon_1 = \frac{1}{3\alpha}\left( \frac{1}{12H^2} - \beta\right).
\end{equation}
Using equations (\ref{f_R}) and (\ref{H_Inf}), the additional slow-roll parameters become
\begin{equation}
\varepsilon_3 = - \frac{24\alpha H^2}{1+24\alpha H^2}  \ \varepsilon_1 
\end{equation}
and 
\begin{equation}
\varepsilon_4 = - \left( \varepsilon_1 + 2 \eta + 2 \frac{\beta}{\alpha} \varepsilon_1\right) 
\end{equation}
where $\eta = -\frac{\ddot{H}}{2H\dot{H}}$ is the second slow-roll parameter in the Hubble rate. 

Further, the derivative of the slow-roll parameters give
\begin{equation}
\dot{\varepsilon}_1 = \left(\frac{2}{1-12\beta H^2}\right)H \varepsilon_1^2 = \frac{\varepsilon_1}{18\alpha H}
\end{equation}
\begin{equation}
\dot{\varepsilon}_3 = - \frac{24\alpha H^2}{1+24\alpha H^2}  \ \dot{\varepsilon}_1 -\left(\frac{2}{1+24\alpha H^2}\right)\frac{\dot{H}}{H} \varepsilon_3 
\end{equation}
so that 
\begin{equation}
\dot{\varepsilon_3} \approx -  \dot{\varepsilon_1}
\end{equation}
since $\alpha H^2\gg 1$ during inflation. The time derivative of the remaining slow-roll parameter turns out to be
\begin{equation}
\dot{\varepsilon}_4 = - \left( \dot{\varepsilon}_1 + 2 \dot{\eta} + 2 \frac{\beta}{\alpha} \dot{\varepsilon}_1\right) 
\end{equation}

We see that $\dot{\varepsilon}_i\ll\varepsilon_i \ H$ since $\alpha H^2\gg 1$ in the begining of inflation. We shall thus take $\dot{\varepsilon}_i \approx 0$ in this regime.

With the above approximation, the scalar spectral index $n_s$ and tensor-to-scalar ratio $r$ in $f(R)$ gravity are given by \cite{Noh2001,Oikonomou2018}
\begin{equation}\label{ns_Equation}
n_s-1 = 3 -2\sqrt{\frac{1}{4} + \frac{(1+\varepsilon_1 - \varepsilon_3 + \varepsilon_4)(2-\varepsilon_3+\varepsilon_4)}{(1-\varepsilon_3)^2} } 
\end{equation}
and
\begin{equation}\label{r_Equation}
r = \frac{48 \varepsilon_3^2}{(1-\varepsilon_3)^2}
\end{equation}

We note that the scalar spectral index $n_s$ and tensor-to-scalar ratio $r$ are functions of $\varepsilon_i$'s, which in turn depend on the values of the parameters $\alpha$ and $\beta$. The value of the parameter $\alpha$ is constrained by the scalaron mass obtained by fitting observed amplitude of the power spectrum, leading to $M \approx 7.5 \times 10^{-4} N_e^{-1} M_{\rm P}$ \cite{Appleby2010}, where $M_{\rm P}$ is the reduced Planck mass, giving
\begin{equation}
\alpha = \frac{1}{6M^2} = \frac{16\pi}{675} N_e^2 \times 10^{8} \ \ell_{\rm P}^2 = 2.68083\times 10^{10} \ \ell_{\rm P}^2 
\end{equation}
for $N_e =60$. Employing this value of $\alpha$, we shall fix the parameter $\beta$ using the the observed values of scalar spectral index $n_s$ and the tensor-to-scalar ratio $r$.  

\begin{figure}[h]
\centering
\includegraphics[width=\textwidth]{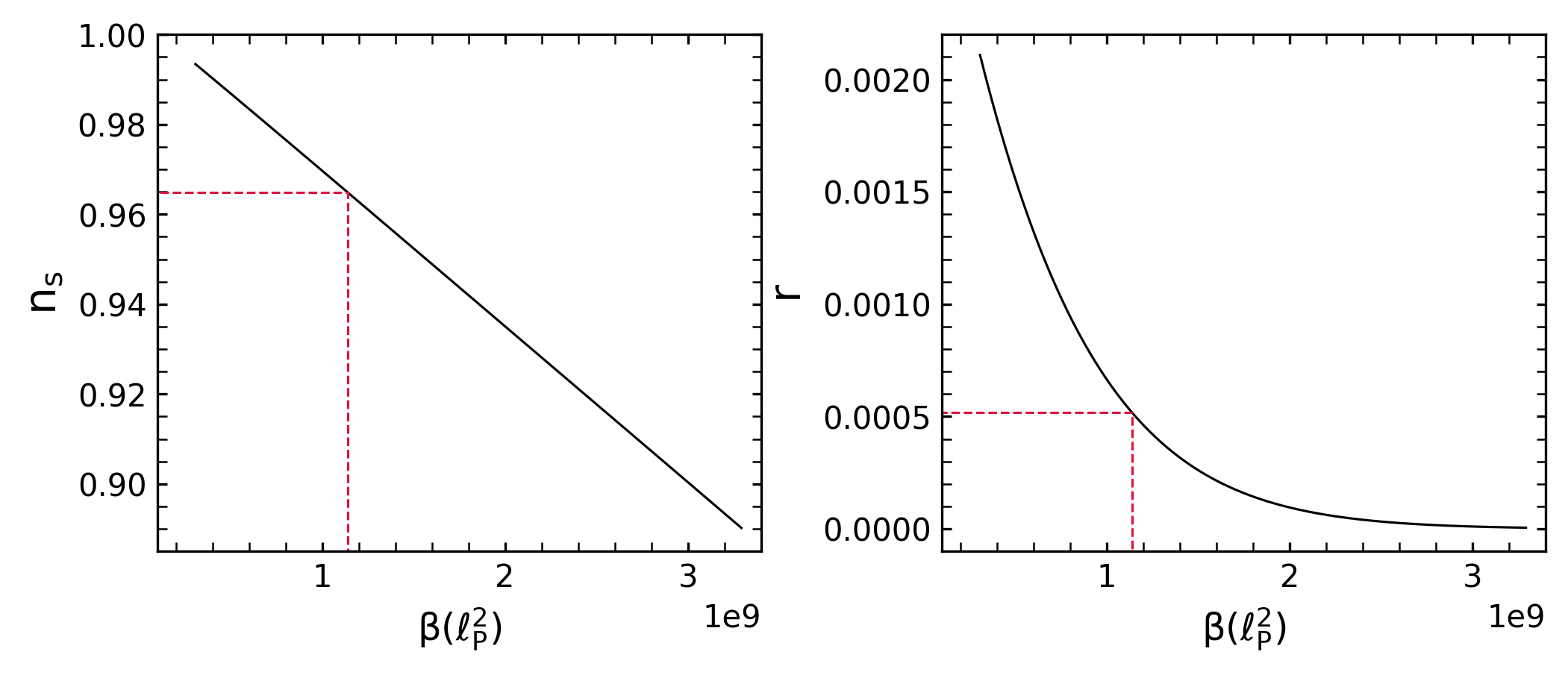}
\caption{(a) Spectral index $n_s$ (left panel) with respect to the scalaron parameter $\beta$ following from equation (\ref{ns_Equation}) where the slow-roll parameters are obtained from the exact numerical integration of equations (\ref{H_Equation}) and  (\ref{rho_D_Equation}). The horizontal dashed line indicates the observed mean value $n_s = 0.9649\pm0.0042$ (error bar not shown). The intersection point  $\beta = 1.143\times 10^9 \ \ell_{\rm P}^2$. (b) Tensor-to-scalar ratio $r$ (right panel) with respect to the scalaron parameter $\beta$ following from equation (\ref{r_Equation}) where the slow-roll parameters are obtained from the exact numerical integration of equations (\ref{H_Equation}) and (\ref{rho_D_Equation}). The vertical line, corresponding to $\beta = 1.143\times 10^9 \ \ell_{\rm P}^2$,  intersects the curve at $r=5.1662\times 10^{-4}$.}
\label{Fig_Spectral_Index}
\end{figure}

Since there are tight bounds on the observed values of the inflationary parameters $n_s$ and $r$, we shall perform exact numerical integration of equation (\ref{H_Equation}) coupled with equation (\ref{rho_D_Equation}) so as to obtain accurate numerical values. In order to obtain an acceptable value of $\beta$ consistent with the observed values of $n_s$ and $r$, we carry out the integration in the inflationary phase with proper initial conditions (described below) for a range of $\beta$ values. The results are shown in Figure \ref{Fig_Spectral_Index}.

In this numerical integration, we use the initial condition $H_0$ from equation (\ref{H_0}) with $N_e= 60$ and the second initial condition, $\dot{H}(0)$ from equation (\ref{H_Diff_Equation_Inf}), to begin the quasi-de-Sitter phase with an initial energy density $\rho(0)=0$. The integration is carried out till the end of inflation, which is identified by the slow-roll condition $\varepsilon_1=1$. This numerical integration further yields the duration of inflation, $t_e-t_0 = 1.0204\times 10^7 t_P$.

Figure \ref{Fig_Spectral_Index} (left panel) shows the variation of $n_s$ with respect to the parameter $\beta$. The horizontal dashed line indicates the observed value of $n_s= 0.9649\pm0.0042$ (without the error bar) from the recent Planck 2018 data \cite{Akrami2020}.  The intersection point gives
\begin{equation}
\beta = 1.143\times 10^9 \ \ell_{\rm P}^2.
\end{equation}

Figure \ref{Fig_Spectral_Index} (right panel) shows the tensor-to-scalar ratio $r$ with respect to $\beta$ resulting from the same numerical intergration. The vertical dashed line at the Planck value of $\beta = 1.143\times 10^9 \ \ell_{\rm P}^2$ intersects the curve at $r=5.1662\times 10^{-4}$. This value of $r$ is well within the observed bound of $r<0.056$ \cite{Akrami2020}.

\section{Analysis of the Reheating Phase}\label{Reh_analysis} 

In the previous section, we saw that the inflationary phase given by equation (\ref{H_Inf}) ends as $t$ approaches $t_e$ and the Hubble parameter falls to a very small value $H_e$. At this time, the growth in density given by equation (\ref{rho_e}) is insignificant. Comparing the expansion term $4H\rho$ and the source term $\lambda$ in the density equation (\ref{rho_D_Equation}), we see that $4H_e\rho_e\ll \lambda$, as $H_e\ll H_S$. Since the expansion term is negligible compared to the source term, it gives the right condition for growth in density implying the beginning of reheating at a time $t_i\sim t_e$.

The Hubble parameter is small during the reheating phase, and initially $H\sim H_e$. Moreover, since $\beta\ll \alpha$, the field equation (\ref{H_Equation}) reduces to 
\begin{equation}\label{H_r}
2H \ddot{H} + 6 H^2 \dot{H} - \dot{H}^2  + \frac{H^2}{6\alpha}    = \frac{\kappa}{18\alpha} \rho, 
\end{equation}
for $t>t_i$, where $t_i$ marks the initial time of the reheating phase.

\subsection{Beginning of Reheating}

At the beginning of reheating, the initial density is negligible, $\rho_i\sim\rho_e\approx0$. We can therefore write
\begin{equation}\label{H_Equation_BR}
2H \ddot{H} + 6 H^2 \dot{H} - \dot{H}^2  + \frac{H^2}{6\alpha}    \approx 0, 
\end{equation}
in the beginning of reheating. 

As the Hubble parameter is small, the term $6H^2\dot{H}$ is smaller compared to the other terms. Neglecting this term, we write
\begin{equation}\label{H_Equation_BR_2}
2H \ddot{H} - \dot{H}^2  + \frac{H^2}{6\alpha}    = 0
\end{equation}
The solution of this equation can be immediately obtained as 
\begin{equation}\label{j}
H(t) =  \psi \cos^2 \left(\frac{t}{2\sqrt{6\alpha}}+\phi \right),
\end{equation}
where $\psi$ represents the amplitude and $\phi$ the initial phase.  The phase of the above oscillation can be fixed by fixing the initial value of the Hubble parameter as $H(t_i)=H_i$ at the initial time $t=t_i$ of the reheating phase. Substituting this initial condition in equation (\ref{j}), we have  $H_i = \psi \cos^2 (t_i/2\sqrt{6\alpha}+\phi ) $. Now, if we set the initial phase $\phi=-t_i/2\sqrt{6\alpha}$, we obtain
\begin{equation}
H(t) = \psi \cos^2 \left(\frac{t-t_i}{2\sqrt{6\alpha}}\right).
\end{equation}

This expression obviously implies a Universe oscillating forever. This is an artefact of neglecting the term $6 H^2 \dot{H}$, whose contribution in determining the correct behaviour of the expansion rate is thus crucial. The effect of the neglected term can be obtained by promoting the amplitude $\psi$ to a function of time, so that 
\begin{equation}\label{cos2}
H(t)  = \psi(t) \cos^2 \left(\frac{t-t_i}{2\sqrt{6\alpha}}\right) = \psi(t) \mu(t).
\end{equation}
Since $6 H^2 \dot{H}$ is small, we expect the amplitude $\psi(t)$ to be a slowly varying function of time. 

Substituting $\psi(t) \mu(t)$ for $H$ in equation (\ref{H_Equation_BR}), we obtain
\begin{equation}
\mu \left(2 \ddot{\psi}  -\frac{\dot{\psi}^2}{\psi} \right)   + 2\dot{\mu} \dot{\psi} + 6\mu \psi \left( \mu\dot{\psi} + \dot{\mu}\psi \right)  = 0. 
\end{equation}
Neglecting the small contribution from the terms $\dot{\psi}^2/\psi$, we obtain the solution 
\begin{equation}\label{psi}
\psi(t) = 4\left[3t - 4 c_2+ 3\sqrt{6\alpha}\sin \left( \frac{t-t_i}{\sqrt{6\alpha}}\right) \right]^{-1}.
\end{equation}

This leads to the  Hubble rate 
\begin{equation}
H(t) = \frac{4}{3}\frac{\cos^2 \left(\frac{t-t_i}{2\sqrt{6\alpha}}\right)}{t - \frac{4 c_2}{3} + \sqrt{6\alpha}\sin \left( \frac{t-t_i}{\sqrt{6\alpha}}\right) },
\end{equation}
in the beginning of reheating phase.

The integration constant $c_2$ can be fixed from the initial value $H(t_i)=H_i$, giving $4c_2 = 3t_i - 4/H_i$. Thus the Hubble's parameter takes the form
\begin{equation}\label{H_BR}
H(t) = \frac{4}{3}\frac{\cos^2 \left(\frac{t-t_i}{2\sqrt{6\alpha}}\right) } 
{(t -t_i) + \frac{4}{3H_i} + \sqrt{6\alpha}\sin \left( \frac{t-t_i}{\sqrt{6\alpha}}\right) }
\end{equation}
in the beginning of reheating, $t\gtrsim t_i$. It is important to note that the amplitude of the Hubble rate varies approximately as $C+(t-t_i)^{-1}$, implying a damped oscillation in this region. 

Using the above expression for the Hubble rate, the scalar curvature $R$ is obtained as $R= -\sqrt{6/\alpha} H \tan[(t-t_i)/2\sqrt{6\alpha}] +3  H^2$. Since the first term dominates, we can approximate this as 
\begin{equation}\label{R_BR}
R(t) = \sqrt{\frac{8}{3\alpha}} \frac{ \sin \left(\frac{t-t_i}{\sqrt{6\alpha}} +\pi \right)}
{ (t -t_i) + \frac{4}{3H_i} + \sqrt{6\alpha}\sin \left( \frac{t-t_i}{\sqrt{6\alpha}}\right) }
\end{equation}
Note that the frequency of oscillation $1/\sqrt{6\alpha}$ is twice that of the Hubble's rate. 

\subsection{Estimate for the End of Reheating}

As we shall show in section \ref{intermediate}, there is negligible back reaction on the curvature scalar due to particle creation in the reheating phase. Consequently, an approximate estimate for the end of reheating occurring after a sufficiently long time $t\sim t_r$ can be obtained from the expression (\ref{R_BR}). From the sufficiency condition (\ref{condition}), a cross-over happens when the left-hand side is approximately equal to the right-hand side, that is 
\begin{equation}\label{end_condition}
 \alpha \langle R^2 \rangle \approx  \frac{128\pi}{3\sqrt{3}}\frac{\varepsilon^4}{m_{\rm P}^2}, 
\end{equation}
where we have taken average quantities, since the time period of oscillation $\tau_{\rm osc}=2\pi\sqrt{6\alpha}$ is expected to be much smaller than the time span $t-t_i$. 

During the phase $t_i\lesssim t\lesssim t_r$, the scalar curvature goes approximately as 
\begin{equation}
R\sim \sqrt{\frac{8}{3\alpha}} \frac{1}{t-t_i}\sin \left(\frac{t-t_i}{\sqrt{6\alpha}}\right),
\end{equation}
and the above condition (\ref{end_condition}) is satisfied at a later time $t_r$, leading to
\begin{equation}
\frac{1}{t_r-t_i} \approx  \sqrt{\frac{32\pi}{3^{1/2}}}\frac{\varepsilon^2}{m_{\rm P}}
\end{equation}
or 
\begin{equation}\label{t_r}
t_r-t_i \approx  \sqrt{\frac{3^{1/2}}{32\pi}} \left(\frac{m_{\rm P}}{\varepsilon}\right)^2 t_{\rm P}. 
\end{equation}

For particle energy $\varepsilon=M/2 = 1.2467\times10^{-6}$ m$_{\rm P} = 1.5221 \times 10^{13}\sim 10^{13}$ GeV, we thus obtain
\begin{equation}
t_r-t_i \approx 8.44513\times 10^{10} \, t_{\rm P} \sim 10^{11} \, t_{\rm P},
\end{equation}
which is much larger than the time period $\tau_{\rm osc} = 2.52 \times 10^6 \, t_{\rm P} \sim10^6 \, t_{\rm P}$, justifying the use of the mean square value $\langle R^2 \rangle$ in the sufficiency condition.

\subsection{Energy Density in the Beginning of Reheating}

Since the sufficiency condition (\ref{condition}) is satisfied in the initial oscillations for $t>t_i$, particle production occurs at a constant rate $\lambda=8\Gamma \varepsilon^4/3\sqrt{3}$.  Substituting the Hubble rate from equation (\ref{H_BR}), the integral $u(t)$, given by (\ref{J}), is expressed as 
\begin{equation}\label{}\small
u(t) =  \frac{8\Gamma \varepsilon^4}{3\sqrt{3}}\int_{t_i}^t dt' \exp \left\{ 4(1+\omega)\int_{t_i}^{t'} dt''   \frac{\cos^2 \left(\frac{t''-t_i}{2\sqrt{6\alpha}}\right) }{(t'' -t_i) + \frac{4}{3H_i} + \sqrt{6\alpha}\sin \left( \frac{t''-t_i}{\sqrt{6\alpha}}\right)} \right\}.
\end{equation}
Performing the inner integration, this reduces to 
\begin{equation}\small\label{}
u(t) = \left(\frac{3H_i}{4}\right)^{2(1+\omega)} \left(\frac{8}{3\sqrt{3}}\right)\Gamma \varepsilon^4
  \int_{t_i}^t     \left\{ (t' -t_i ) + \frac{4}{3H_i} + \sqrt{6\alpha}\sin \left(\frac{t'- t_i}{\sqrt{6\alpha}}\right) \right\}^{2(1+\omega)}  dt'.
\end{equation}
The first term in the above integrand is the only growing term compared to the other terms.  Further, we see that the second and third terms are of the same order.  Since the last term is oscillatory about zero, we shall drop its effect while taking the area under the curve. In this approximation, we obtain
\begin{equation}\label{}
u(t) \approx  \left(\frac{3H_i}{4}\right)^{2\omega+2}    \left(\frac{8}{3\sqrt{3}}\right) \frac{\Gamma \varepsilon^4}{2\omega+3} \left\{  t -t_i  + \frac{4}{3H_i} \right\}^{2\omega+3}
\end{equation}
Since the evolution of density in any region is given by the expression (\ref{rho_Equation}), we obtain 
\begin{equation}\small \label{rho_IR}
\rho(t) \approx  \frac{8}{11\sqrt{3}} \Gamma \varepsilon^4  \left\{ t-t_i + \frac{4}{3H_i} \right\}
\end{equation}
In obtaining the above expression from (\ref{rho_Equation}), we dropped the effect of the initial density $\rho_0$, as the exponent $\exp{\left\{ -4\int H dt'\right\}}$ makes this term insignificant as the Universe expands. On the other hand, there is a significant effective growth in density coming from $u(t)$ although the exponential factor attenuates with the expansion of the Universe. Thus to a good approximation, we find that the density grows linearly in this region.

\subsection{Intermediate Region of the Reheating Phase}\label{intermediate}

The back-reaction of the particle creation on the background metric can be analyzed by considering the trace equation (\ref{modi_trace_equation}) with the production rate given by equation (\ref{const_condition}). Since the scalar curvature $R\ll\beta^{-1}$ during reheating, we shall drop the quadratic term giving 
\begin{equation}
\ddot{R}+3H\dot{R} + \frac{R}{6\alpha}=   \frac{4}{9\sqrt{3}}\frac{\kappa}{\alpha} \frac{\Gamma \varepsilon^4}{H}. 
\end{equation}
The above equation represents a forced damped oscillator with a time dependent damping coefficient $3H(t)$ and and a force inversely proportional to the Hubble rate. To solve this differential equation, the time average of the Hubble parameter obtained in the initial phase of reheating given by equation (\ref{H_BR}) can be employed. Since the amplitude of the Hubble rate is slowly varying compared to the rapidly oscillating term, its time average can be approximated as $\langle H\rangle\approx \frac{2}{3} (t-t_i)^{-1}$, leading to 
\begin{equation}\small
\ddot{R}+\frac{2}{(t-t_i)}\dot{R} + \frac{R}{6\alpha} = \frac{128\pi}{3\sqrt{3}} \frac{\Gamma}{4\alpha}   \frac{\varepsilon^4}{m_{\rm P}^2} (t-t_i). 
\end{equation}
This gives the solution
\begin{equation}\small
R = \frac{64\pi}{\sqrt{3}} \frac{\Gamma\varepsilon^4}{m_{\rm P}^2} \left\{ (t-t_i) -\frac{12\alpha}{(t-t_i)}\right\} + A \frac{\sin \left(\frac{t-t_i}{\sqrt{6\alpha}}+\phi \right)}{t-t_i}.
\end{equation}
The first term proportional to $\Gamma$ represents the effect of back-reaction whereas the second term with the integration constant $A$ represents the behaviour in the absence of back-reaction. Soon after few oscillations, the term $t-t_i$ becomes larger than $12\alpha (t-t_i)^{-1}$. Moreover, in the absence of back-reaction,  the oscillatory term should represent an approximation for the expression given by (\ref{R_BR}). Thus, to a good approximation, the integration constant $A$ must coincide with $\sqrt{8/3\alpha}$, leading to 
\begin{equation}\small
R \approx \frac{64\pi}{\sqrt{3}}\frac{\Gamma\varepsilon^4}{m_{\rm P}^2} (t-t_i)  + \sqrt{\frac{8}{3\alpha}} \frac{\sin \left(\frac{t-t_i}{\sqrt{6\alpha}}+\phi \right)}{t-t_i}.
\end{equation}

Thus, the back-reaction of particle production on the scalar curvature becomes important when 
\begin{equation}
t_{rb}-t_i \gtrsim \frac{3^{1/4}}{2\sqrt{2}} \left(\frac{\Gamma^{-1}}{\tau_{\rm osc}}\right)^{1/2} \left(\frac{m_P}{\varepsilon}\right)^2 \ t_{\rm P}
\end{equation}

An estimate for $\Gamma$ can be obtained from the uncertainty relation $\Delta E \, \Delta t\sim \hbar/2$ for pair production via the decay of a scalaron of mass $M$. This gives $2\varepsilon \,\Gamma^{-1}\sim \hbar/2$, so that $\Gamma\sim4\varepsilon$. Thus  $\Gamma^{-1} \sim \frac{m_{\rm P}}{4\varepsilon} \, t_{\rm P} = 3.1168 \times 10^{5} \, t_{\rm P}$ for $\varepsilon = M/2 = 1.2467 \times 10^{-6}$ m$_P$. Since $\Gamma^{-1}/\tau_{\rm osc}\sim 10^{-1}$, the particles are created at a rate faster than the rate of oscillation. This justifies using the average energy $\varepsilon$ per particle in our estimates. Moreover, since $t_{rb}-t_i = 1.0529\times 10^{11} t_{\rm P} > t_r-t_i$, the back-reaction becomes important after the end of reheating. Thus, the back-reaction on the curvature field is insignificant during reheating. 

The Hubble rate in this region can be obtained by using equation (\ref{rho_IR}) in (\ref{H_r}), yielding 
\begin{equation}\label{Intermediate_H}\small
2H \ddot{H} + 6 H^2 \dot{H} - \dot{H}^2  + \frac{H^2}{6\alpha}   =  \frac{4}{99\sqrt{3}}\frac{\Gamma}{\alpha}  \frac{\varepsilon^4}{m_{\rm P}^2} (t-t_i).
\end{equation}

For vanishing right-hand side, the solution is given by equation (\ref{H_BR}) which has the form $H(t) = \mu(t)\psi(t)$. As a solution of equation (\ref{Intermediate_H}), we shall assume an approximate form $H(t) = \psi(t)\mu(t)+\phi(t)$, with the same functional forms for $\psi(t)$ and $\mu(t)$ as before. Consequently, we seek a solution for the unknown function $\phi(t)$. Substituting in equation (\ref{Intermediate_H}), we obtain a differential equation for $\phi(t)$, given by 
\begin{multline}\label{}
2\mu^2\psi\ddot{\psi} + 2\mu\dot{\mu}\psi\dot{\psi}  + \psi^2\left( 2\mu\ddot{\mu} - \dot{\mu}^2 + \frac{\mu^2}{6\alpha}\right) + 2\mu\psi\ddot{\phi} +6\mu^3\psi^2\dot{\psi}+ 6\mu^2\dot{\mu}\psi^3  -\dot{\psi}^2\mu^2 \\
+ 2(\mu\ddot{\psi} +2\dot{\psi}\dot{\mu}  + \psi\ddot{\mu})\phi + 2\phi \ddot{\phi} 
+6 (2 \psi\mu\phi+\phi^2)(\dot{\psi}\mu + \psi\dot{\mu}) +6(\psi^2\mu^2 + 2 \psi\mu\phi+\phi^2) \dot{\phi} \\
-( 2\mu\dot{\psi}\dot{\phi} + 2\dot{\mu}\psi\dot{\phi} + \dot{\phi}^2) 
+ \frac{1}{6\alpha} (2 \mu\psi\phi+\phi^2) = \frac{4}{99\sqrt{3}} \frac{\Gamma}{\alpha}  \frac{\varepsilon^4}{m_{\rm P}^2} (t-t_i)
\end{multline}
Neglecting the term $\mu^2\dot{\psi}^2$, and using equation (\ref{cos2}) and (\ref{psi}),  a few terms in the above equation cancel out, reducing it to \begin{multline}\label{after_removing_small_time_scales}
2(\mu\ddot{\psi} +2\dot{\psi}\dot{\mu}  + \psi\ddot{\mu})\phi + 2\phi \ddot{\phi} +6 (2 \psi\mu\phi+\phi^2)(\dot{\psi}\mu + \psi\dot{\mu}) +6(\psi^2\mu^2 + 2 \psi\mu\phi+\phi^2) \dot{\phi}\\
-( 2\mu\dot{\psi}\dot{\phi} + 2\dot{\mu}\psi\dot{\phi} + \dot{\phi}^2) 
+ \frac{1}{6\alpha} (2 \mu\psi\phi+\phi^2) = \frac{4}{99\sqrt{3}} \frac{\Gamma}{\alpha}  \frac{\varepsilon^4}{m_{\rm P}^2} (t-t_i)
\end{multline}

Since $\dot{\mu} = - \frac{1}{2\sqrt{6\alpha}} \sin\left( \frac{t-t_i}{\sqrt{6\alpha}} \right)$  and  $\ddot{\mu}=\frac{1}{12\alpha}(1-2\mu)$, we have $\mu\propto (\tau_{\rm osc})^0$, $\dot{\mu}\propto (\tau_{\rm osc})^{-1}$, $\ddot{\mu}\propto (\tau_{\rm osc})^{-2}$. Since the right hand side involves a larger time scale, the average evolution of Hubble's rate is affected only in such time scales. Therefore we shall neglect the terms that involve smaller time scales, namely $\tau_{\rm osc}$. Moreover, these terms average out to zero since they are rapidly oscillating. Thus equation (\ref{after_removing_small_time_scales}) reduces to 
\begin{multline}\label{}
2\mu\ddot{\psi}\phi + 2\phi \ddot{\phi} +6 (2 \psi\mu\phi+\phi^2)\dot{\psi}\mu   -2\mu\dot{\psi}\dot{\phi}    - \dot{\phi}^2   +6(\psi^2\mu^2 + 2 \psi\mu\phi+\phi^2) \dot{\phi} =  \frac{4}{99\sqrt{3}} \frac{\Gamma}{\alpha}  \frac{\varepsilon^4}{m_{\rm P}^2} (t-t_i)
\end{multline}

Substituting the time-averaged values of $\mu=\frac{1}{2}$ and $\psi=\frac{4}{3}\frac{1}{t-t_i}$, and assuming a power-like trial solution, $\phi=B (t-t_i)^\nu$, we obtain
\begin{multline}\small\label{}
\frac{4}{3}\bigg\{ 2B - 4 B +  2\nu B  +\nu B  \bigg\} (t-t_i)^{\nu-3} 
+ \bigg\{ 2\nu(\nu-1)B^2 - 4B^2 + 8\nu B^2 - \nu^2 B^2  \bigg\} (t-t_i)^{2\nu-2} \\
 +6\nu B^3   (t-t_i)^{3\nu-1}   = \frac{4}{99\sqrt{3}} \frac{\Gamma}{\alpha} \frac{\varepsilon^4}{m_{\rm P}^2} (t-t_i)
\end{multline}

For $\nu\leq0$, we observe that the powers on both sides of the above equation cannot be equated. On the other hand, for $\nu>0$, it is possible to equate the powers on both sides. For large times $t-t_i$, the last term on the left-hand side dominates. Thus, equating powers on both sides, we obtain 
\begin{equation}
3\nu -1 = 1 \implies \nu = \frac{2}{3}, 
\end{equation}
and equating the coefficients, we have 
\begin{equation}\label{B}
B   = \left( \frac{1}{99\sqrt{3}}\frac{\Gamma}{\alpha}  \frac{\varepsilon^4}{m_{\rm P}^2}  \right)^{1/3}. 
\end{equation}
Consequently, an approximate solution in the intermediate region has the form 
\begin{equation}
H(t) = \psi(t) \, \mu(t) + \left( \frac{\Gamma}{99\sqrt3{}\alpha} \,  \frac{\varepsilon^4}{m_{\rm P}^2}  \right)^{1/3} (t-t_i)^{2/3}. 
\end{equation}
The effect of back reaction thus gives rise to an increasing component (on an average) in the Hubble rate. We have previously seen a similar effect of back-reaction in the scalar curvature that goes like $t-t_i$. 

The effect of back-reaction on the Hubble rate becomes important when 
\begin{equation}
t_{hb}-t_i \geq \left\{ \frac{11}{3\sqrt{3} \pi^2} \left( \frac{\Gamma^{-1} \tau_{osc}^2 }{t_{\rm P}^3}\right) \left(\frac{m_{\rm P}}{\varepsilon}\right)^4 \right\}^{1/5} \, t_{\rm P}
\end{equation}
or
\begin{equation}
t_{hb}-t_i \geq 1.7741 \times 10^8 \, t_{\rm P}.
\end{equation}
Since $t_r-t_i\sim10^{11}\, t_{\rm P}$, the Hubble rate encounters the effect of back-reaction within the reheating phase.

In this region, the density equation can be written as
\begin{equation}
\dot{\rho} + 4\left\{ \frac{2}{3}\frac{1}{t-t_i} + B (t-t_i)^{2/3}  \right\}\rho = \frac{8}{3\sqrt{3}}\Gamma \varepsilon^4
\end{equation}
giving the solution
\begin{equation}\small
\rho(t) = \frac{8}{3\sqrt{3}}\Gamma \varepsilon^4 \frac{e^{-\frac{12}{5}B (t-t_i)^{5/3}}}{(t-t_i)^{8/3}} \int_{t_i}^t (t'-t_i)^{8/3} e^{\frac{12}{5}B (t'-t_i)^{5/3}} dt' 
\end{equation}

For $t\gtrsim t_i$, that is very close to the beginning of reheating, the above integration can be evaluated by substituting unity for the exponential so that $\rho(t)\approx\frac{8}{11\sqrt{3}}\Gamma \varepsilon^4 (t-t_i)$ which corresponds well with equation (\ref{rho_IR}) derived earlier in the beginning of reheating. The slight mismatch in the initial time is due to the approximations involved in deriving the above equation in the intermediate region. 

For $t\gg t_i$,  that is, in the intermediate region, we can evaluate the above integral by estimating the value of $B$ obtained from equation (\ref{B}), by writing it as 
\begin{equation}
B   = \left\{ \frac{8\pi^2}{33\sqrt{3}}\frac{t_{\rm P}^2}{\Gamma^{-1}\tau_{\rm osc}^2}  \left(\frac{\varepsilon}{m_{\rm P}}\right)^4  \right\}^{1/3}  \sim 1.7386 \times 10^{-14} \ t_{\rm P}^{-5/3}.
\end{equation}
Taking $t-t_i\sim 10^7 \, t_{\rm P}$, we find that $\frac{12}{5}B (t-t_i)^{5/3}\sim10^{-2}$, so that the exponent is a very small number. Consequently, for $t\gg t_i$, we can expand the exponential inside the integral, giving the approximate solution
\begin{equation}\small
\rho(t) \approx \frac{8}{11\sqrt{3}} \Gamma \varepsilon^4  (t-t_i)e^{-\frac{12}{5}B (t-t_i)^{5/3}} \left\{ 1+  \frac{33}{20}B (t-t_i)^{5/3}\right\}.
\end{equation}
This indicates that the density declines from the linear growth for $t\gg t_i$, that is in the intermediate region of the reheating phase. 

\subsection{End of Reheating}

According to the present scenario, at $t_r$ given by equation (\ref{t_r}), the sufficiency condition (\ref{condition}) reaches its minimum so that $t_r$ can be identified as the time when reheating ends. At subsequent times $t>t_r$, the available energy in the scalaron field becomes insufficient to populate all available cells in the configuration space and hence the energy density of the Universe starts declining. Thus $\rho(t)$ will have a maxima at time $t_r$ when $\dot{\rho}(t_r)=0$. Consequently, the density equation (\ref{rho_D_Equation}) gives the maxima as 
\begin{equation}
\rho_r = \frac{2}{3\sqrt{3}}\frac{\Gamma}{H_r}\varepsilon^4,
\end{equation}
where $H_r$ is the average Hubble rate at the end of reheating. Since the reheating temperature $T_r$ can be identified as $\rho_r = \frac{\pi^2}{30} g^* T_r^4$, this immediately yields
\begin{equation}
T_r = \left(\frac{20}{\sqrt{3}\pi^2 } \frac{\Gamma}{g^*H_r} \right)^{1/4} \varepsilon, 
\end{equation}
where $g^*$ is the total number of degrees of freedom. For energies above 1 TeV, all degrees of freedom in the Standard Model are relativistic, resulting in $g^* = 106.75$ \cite{D'Eramo_2017}. Since the end of reheating can be identified with the Hubble rate $H$ approaching the value of particle creation rate $\Gamma$ ($H\approx\Gamma$),  the reheating temperature can be estimated as 
\begin{equation}
T_r = 4.9248 \times 10^{13} \ {\rm GeV}. 
\end{equation}

To this end, we shall analyse the period $t\lesssim t_r$, that is, as the end of reheating is approached. We suppose a solution of the form $H(t)= \Gamma + \chi(t)$ in the region $t\lesssim t_r$, where $\chi \ll \Gamma$. Using the change of variable $\xi = t_r-t$, equation (\ref{H_r}) reduces to 
\begin{equation}\label{}
 \chi'' - 3 \Gamma \chi'  +  \frac{1}{6\alpha}  \chi    =  \frac{\Gamma}{12\alpha}  \left( \frac{\pi}{12\sqrt{3}} \frac{M^2}{m_{\rm P}^2}   -   1\right),
\end{equation}
up to linear order in $\chi(\xi)$. Here the primes denote differentiation with respect to $\xi$. Since $M\ll m_{\rm P}$, this equation can be approximated as 
\begin{equation}\label{}
 \chi'' - 3 \Gamma \chi'  +  \frac{1}{6\alpha}  \chi    \approx -\frac{\Gamma}{12\alpha}, 
\end{equation}
giving the solution 
\begin{equation}\small\label{}
 \chi(\xi) = - \frac{\Gamma}{2} + e^{\frac{3\Gamma\xi}{2}} \left\{ C_1 e^{\sigma\xi} + C_2  e^{-\sigma\xi} \right\}
\end{equation}
where $\sigma= \sqrt{\frac{9}{4\Gamma^{-2}}- \frac{1}{6\alpha}}$. Since $\Gamma = 4\varepsilon = 2M$ and using $\alpha = \frac{1}{6M^2}$, we have $\sigma = 2\sqrt{2}M$. This leads to 
\begin{equation}\small\label{}
 \chi(\xi) = - \frac{\Gamma}{2} + e^{\frac{3\Gamma\xi}{2}} \left\{ C_1 e^{2\sqrt{2}M\xi} + C_2  e^{-2\sqrt{2}M\xi} \right\}
\end{equation}
so that the Hubble rate becomes 
\begin{equation}\label{}
H(t) =  \frac{\Gamma}{2} + e^{\frac{3\Gamma}{2}(t_r-t)} \left\{ C_1 e^{2\sqrt{2}M(t_r-t)} + C_2  e^{-2\sqrt{2}M(t_r-t)} \right\}
\end{equation}
Using the condition $H(t=t_r) = \Gamma$, the unknowns are related by $C_1 + C_2 =\frac{\Gamma}{2 }$, leading to 
\begin{equation}\label{H_ER}
H(t) = \frac{\Gamma}{2} \left\{ 1 + e^{\frac{3\Gamma}{2}(t_r-t)}  \right\} -2C_2  e^{\frac{3\Gamma}{2}(t_r-t)} \sinh [2\sqrt{2}M(t_r-t)], 
\end{equation}
in the region $t\lesssim t_r$. 

From the equation $R =6 (\dot{H} + 2H^2)$, we thus obtain 
\begin{multline}\label{R_ER}
R(t) = e^{\frac{3\Gamma}{2}(t_r-t)} \left\{ 18 C_2 \Gamma \sinh 2\sqrt{2}M(t_r-t) + 24\sqrt{2}C_2 M \cosh 2\sqrt{2}M(t_r-t) -\frac{9}{2}\Gamma^2\right\} \\
 + 48 C_2^2 e^{3\Gamma(t_r-t)} \sinh^2 2\sqrt{2}M(t_r-t) \\
+ 3\Gamma^2 \left( 1+ e^{\frac{3\Gamma}{2}(t_r-t)}\right)^2 -C_2 \Gamma \left( 1+ e^{\frac{3\Gamma}{2}(t_r-t)}\right)  e^{\frac{3\Gamma}{2}(t_r-t)} \sinh 2\sqrt{2}M(t_r-t)
\end{multline}

As we have discussed earlier, the sufficiency condition (\ref{condition}) reaches the equality at $t=t_r$. Employing equation (\ref{R_ER}), this condition leads to  
\begin{equation}\label{}
24\sqrt{2} C_2M + 60M^2 = \sqrt{\frac{16\pi}{\sqrt{3}}} \frac{M^3}{m_P}
\end{equation}
This gives
\begin{equation}
C_2 \approx -\frac{5}{2\sqrt{2}} M
\end{equation}
The Hubble rate given by equation (\ref{H_ER}) is obtained as 
\begin{equation}\label{FH_ER}
H(t) = \frac{\Gamma}{2} \left\{ 1 + e^{\frac{3\Gamma}{2}(t_r-t)}  \right\} +\frac{5M}{\sqrt{2}}  e^{\frac{3\Gamma}{2}(t_r-t)} \sinh 2\sqrt{2}M(t_r-t),
\end{equation}
in the region $t\lesssim t_r$. Thus we see that the Hubble rate approaches $\Gamma$ at $t=t_r$ from a higher value at $t\lesssim t_r$ where this approximate expression is valid. It is important to note that we cannot extrapolate the above relations to regions $t\ll t_r$.

\subsection{Preheating and Thermilization} 

In the previous subsections, we described the details of the dynamical evolution in different regions of the reheating phase. To this end, we identify and analyze the preheating stage with the physical process of thermilization.

The mean free path $\ell_{\rm mean}(t)$ of the particles is related to the energy density $\rho(t)$ as
\begin{equation}
\ell_{\rm mean}(t) \approx \left\{ \frac{3}{4\pi}\frac{\varepsilon}{\rho(t)} \right\}^{1/3}. 
\end{equation}
Since the particles are expected to be ultra-relativistic, an estimate for collision time-scale is given by 
 \begin{equation}
 \tau_{\rm coll} (t) \approx \ell_{\rm mean}(t), 
\end{equation}
and the collision rate $ \Gamma_{\rm coll} (t)$ can be estimated as 
\begin{equation}
 \Gamma_{\rm coll} (t) \approx \frac{1}{\tau_{\rm coll}(t)}. 
 \end{equation}
It is important to note that the collision rate $ \Gamma_{\rm coll} (t)$ is a physically different quantity from the particle creation rate $\Gamma$.

In the beginning of the reheating phase, that is for $t> t_i$, from equation (\ref{rho_IR}) an approximate estimate for the energy density is  $\rho(t)\approx \frac{8}{11\sqrt{3}}\Gamma\varepsilon^4 (t-t_i)$,  so that the collision rate is estimated as 
\begin{equation}
 \Gamma_{\rm coll} (t) \approx \left\{ \frac{32\pi}{33\sqrt{3}} \frac{(t-t_i)}{\Gamma^{-1}} \right\}^{1/3} \varepsilon, 
 \end{equation}
and the average Hubble rate has the behaviour $H(t)\approx \frac{2}{3}\frac{1}{t-t_i}$ in this region. These expressions indicate that $ \Gamma_{\rm coll}(t)\ll H(t)$ for $t\gtrsim t_i$. Consequently, the collision rate cannot catch up with the Hubble expansion rate and the system of particles cannot reach thermal equilibrium. This stage is the preheating phase.

For longer times $t>t_i$, the collision rate $ \Gamma_{\rm coll}(t)$ increases whereas the Hubble rate decreases. At a time $t_{\rm th}$ (say), the collision rate  $ \Gamma_{\rm coll}(t)$ catches up with the Hubble rate $H(t)$. Consequently,  $ \Gamma_{\rm coll}(t_{\rm th})= H(t_{\rm x})$ gives an estimate for the beginning of thermilization at a time $t_{\rm th}$ as 
\begin{equation}
t_{\rm th}-t_i = \left(\frac{11}{12\sqrt{3}\pi}\frac{\Gamma^{-1}}{\varepsilon^3}\right)^{1/4} , 
\end{equation}
from which we obtain an estimate $t_{\rm th}-t_i= 4.0572 \times 10^5 \, t_{\rm P}$. This estimate for $t_{\rm th}-t_i$ is from the region when the back-reaction on the Hubble rate has a negligible effect on the dynamics. As we have seen earlier, the back-reaction becomes effective only in the intermediate region when $t_{hb}-t_i \sim 10^8 \, t_{\rm P}$.

The above estimates for $t_{\rm th}-t_i$ and $t_{hb}-t_i$,  indicates that the thermilization process begins much before the time when the back-reaction on the Hubble rate begins to be effective. At the beginning of thermilization $T = T_{\rm th}$, the temperature is estimate from   
\begin{equation}
T_{\rm th} = \left( \frac{3^{1/8} 80}{11^{3/4}\sqrt{2} } \frac{1}{g^* \pi^{9/2}}   \frac{\varepsilon^{13/4}}{\Gamma^{-3/4}}  \right)^{1/4} 
\end{equation}
so that 
\begin{equation}
T_{\rm th} = 2.8237 \times 10^{12} \ {\rm GeV}. 
\end{equation}
For subsequent times $t>t_{\rm th}$, the condition $ \Gamma_{\rm coll}(t)> H(t)$ holds and the process of thermilization continues.

\section{Conclusion}\label{Conclusion}

In this paper, we considered a physically plausible scenario of reheating following the inflationary phase of a modified gravity model $f(R)=R+\alpha R^2 + \beta R^2 \ln R/\mu^2$ in the Jordan frame. Since particle creation is essentially a quantum mechanical phenomenon, we formulated this scenario based on Heisenberg's uncertainty principle. In addition, we fixed the parameter $\beta$ from the observed spectral index and tensor-to-scalar ratio. 

We find that although particle production happens in the inflationary phase, its contribution to the energy density $\rho$ is negligibly small at the end of inflation owing to the quasi-de-Sitter expansion. On the other hand, particle creation happening in the oscillatory regime following the inflationary phase gives rise to a significant growth in the energy density $\rho$. In the initial stage of this reheating regime, the average density grows linearly with time, whereas the growth deviates from linearity at longer times. Eventually, the reheating period ends when the energy density $\rho$ grows to a maximum value which is marked by the sufficiency condition attaining its minimum value so that the available energy density in the scalaron field equals energy density of the created particles in the available states in the configuration space within a time-span of the inverse decay rate. 

We analyzed the different phases by making analytical estimates for macroscopic quantities. They represent good approximations because of the fact that the frequency of particles creation happens at a faster rate than the oscillation frequency of the Hubble rate so that an average energy $\varepsilon$ per created particles could be used in our estimates. In addition, since the source term in the density equation contains the time scale $\Gamma^{-1}$, the growth in density varies over the same time scale $\Gamma^{-1}$, so that the average Hubble rate yields a good estimate for the growth in energy density. 

In the initial period of reheating, the growth of energy density has a negligible back-reaction on the dynamics of the Hubble rate which goes like $(t-t_i)^{-1}$. At longer time,  in the intermediate stage, the average Hubble rate behaves like $(t-t_i)^{2/3}$, signifying the effect of back-reaction. On the other hand, back-reaction is negligible on the scalar curvature for all times during the entire period of reheating.

In this scenario of reheating, we find a well-defined period of preheating where the particles are unable to reach thermal equilibrium because their collision rate lags behind the Hubble expansion rate. The end of this preheating stage is marked by a time $t_{\rm th}$ determined by the time when the collision rate $\Gamma_{\rm coll}$ catches up with the Hubble expansion rate $H$. We find that the preheating stage spans over a time of $t_{\rm th}-t_i \sim 4.0572 \times 10^{5} \, t_{\rm P}$.  Subsequently, themilization takes place as the collision rate $\Gamma_{\rm coll}$ exceeds the Hubble expansion rate $H$ for $t>t_{\rm th}$. 

As the density keeps on growing beyond the time $t_{\rm th}$, thermilization keeps the system in thermal equilibrium. The reheating period continues so long as the inequality in the sufficiency condition (\ref{condition}) holds,  that is, when sufficient energy is available in the scalaron field to completely populate the configuration space constrained by the Heisenberg uncertainty principle. 

We can analyze the growth in energy density in terms of the average value of $4 H\rho$ which increases so long as the density continues to grow as a result of meeting the sufficiency condition (\ref{condition}). Eventually, $4H\rho$ approaches the value $8\Gamma \varepsilon^4/3\sqrt{3}$ at a long time. This asymptotic approach to a constant value corresponds to the density approaching a maxima $\rho_r$ at a long time $t_r$. At this moment, $\dot{\rho}$ approaches zero in equation (\ref{rho_D_Equation}), since the expansion term compensates for the source term, stopping further growth in density. At time $t_r$, the equality in the sufficiency condition (\ref{condition}) is reached when the available energy is just enough to populate the entire configuration space. Beyond $t_r$, sufficient energy is no longer available to populate the entire configuration space and the energy density $\rho$ starts declining. Thus the density $\rho_r$ is a maxima occurring at time $t_r$, marking the end of the reheating phase. 

Our present scenario, based on Heisenberg's uncertainty principle, facilitated a detailed analysis of all stages of the reheating phase. This includes an analysis of the preheating stage and the subsequent cross-over to the thermilization stage along with a proper identification of the end-of-reheating. It may be fair to say that the present scenario gives a fundamental understanding of the physical processes in the reheating phase although it rests on a few approximate but reasonable physical assumptions.

\acknowledgments

Arun Mathew is indebted to the Indian Institute of Technology Guwahati and  Dublin Institute for Advanced Studies for extending various facilities during his doctoral and post-doctoral programs.



\begin{thebibliography}{10}

\bibitem{Weinberg2008}
S.~Weinberg, \emph{Cosmology}, Oxford University Press, USA (2008).

\bibitem{Peebles1993}
P.J.E.~Peebles, \emph{{Principles of physical cosmology}}, Princeton University
  Press (1993).

\bibitem{Fields1996}
B.D.~Fields and K.A.~Olive, \emph{Model-independent predictions of big bang
  nucleosynthesis},
  \href{https://doi.org/https://doi.org/10.1016/0370-2693(95)01508-6}{\emph{Physics
  Letters B} {\bfseries 368} (1996) 103 }.

\bibitem{Schramm1993}
D.N.~Schramm, \emph{Cosmological implications of light element abundances:
  theory}, \href{https://doi.org/10.1073/pnas.90.11.4782}{\emph{Proceedings of
  the National Academy of Sciences} {\bfseries 90} (1993) 4782}.

\bibitem{Peebles1991}
P.J.E.~Peebles, D.N.~Schramm, E.L.~Turner and R.G.~Kron, \emph{The case for the
  relativistic hot big bang cosmology},
  \href{https://doi.org/10.1038/352769a0}{\emph{Nature} {\bfseries 352} (1991)
  769}.

\bibitem{Mather1994}
J.C.~{Mather}, E.S.~{Cheng}, D.A.~{Cottingham}, J.~{Eplee}, R.~E.,
  D.J.~{Fixsen}, T.~{Hewagama} et~al., \emph{{Measurement of the Cosmic
  Microwave Background Spectrum by the COBE FIRAS Instrument}},
  \href{https://doi.org/10.1086/173574}{\emph{Astrophys. J} {\bfseries 420}
  (1994) 439}.

\bibitem{Tammann1992}
G.A.~Tammann, \emph{Cosmic expansion and deviations from it},
  \href{https://doi.org/10.1088/0031-8949/1992/t43/004}{\emph{Physica Scripta}
  {\bfseries T43} (1992) 31}.

\bibitem{Penzias1965}
A.A.~{Penzias} and R.W.~{Wilson}, \emph{{A Measurement of Excess Antenna
  Temperature at 4080 Mc/s.}},
  \href{https://doi.org/10.1086/148307}{\emph{Astrophys. J.} {\bfseries 142}
  (1965) 419}.

\bibitem{Kolb1990}
E.W.~Kolb and M.S.~Turner, \emph{{The Early Universe}}, vol.~69 (1990).

\bibitem{Guth1981}
A.H.~Guth, \emph{Inflationary universe: A possible solution to the horizon and
  flatness problems},
  \href{https://doi.org/10.1103/PhysRevD.23.347}{\emph{Phys. Rev. D} {\bfseries
  23} (1981) 347}.

\bibitem{LInde1982a}
A.~Linde, \emph{A new inflationary universe scenario: A possible solution of
  the horizon, flatness, homogeneity, isotropy and primordial monopole
  problems},
  \href{https://doi.org/https://doi.org/10.1016/0370-2693(82)91219-9}{\emph{Physics
  Letters B} {\bfseries 108} (1982) 389 }.

\bibitem{LInde1982b}
A.~Linde, \emph{Coleman-weinberg theory and the new inflationary universe
  scenario},
  \href{https://doi.org/https://doi.org/10.1016/0370-2693(82)90086-7}{\emph{Physics
  Letters B} {\bfseries 114} (1982) 431 }.

\bibitem{LInde1982c}
A.~Linde, \emph{Scalar field fluctuations in the expanding universe and the new
  inflationary universe scenario},
  \href{https://doi.org/https://doi.org/10.1016/0370-2693(82)90293-3}{\emph{Physics
  Letters B} {\bfseries 116} (1982) 335 }.

\bibitem{Albrecht1982}
A.~Albrecht and P.J.~Steinhardt, \emph{Cosmology for grand unified theories
  with radiatively induced symmetry breaking},
  \href{https://doi.org/10.1103/PhysRevLett.48.1220}{\emph{Phys. Rev. Lett.}
  {\bfseries 48} (1982) 1220}.

\bibitem{Starobinsky1980}
A.A.~Starobinsky, \emph{A new type of isotropic cosmological models without
  singularity},
  \href{https://doi.org/https://doi.org/10.1016/0370-2693(80)90670-X}{\emph{Physics
  Letters B} {\bfseries 91} (1980) 99 }.

\bibitem{Starobinsky1979}
A.A.~Starobinsky, \emph{Spectrum of relict gravitational radiation and the
  early state of the universe}, {\emph{JETP Lett.} {\bfseries 30} (1979) 682}.

\bibitem{Linde1984}
A.D.~Linde, \emph{The inflationary universe},
  \href{https://doi.org/10.1088/0034-4885/47/8/002}{\emph{Reports on Progress
  in Physics} {\bfseries 47} (1984) 925}.

\bibitem{Linde1983}
A.~Linde, \emph{Chaotic inflation},
  \href{https://doi.org/https://doi.org/10.1016/0370-2693(83)90837-7}{\emph{Physics
  Letters B} {\bfseries 129} (1983) 177 }.

\bibitem{Linde1985}
A.~Linde, \emph{Initial conditions for inflation},
  \href{https://doi.org/https://doi.org/10.1016/0370-2693(85)90923-2}{\emph{Physics
  Letters B} {\bfseries 162} (1985) 281 }.

\bibitem{Dolgov1989}
A.~Dolgov and D.~Kirilova, \emph{Production of particles by a variable scalar
  field}, {\emph{Sov. J. Nucl. Phys.} {\bfseries 51} (1989) 172}.

\bibitem{Traschen1990}
J.H.~Traschen and R.H.~Brandenberger, \emph{Particle production during
  out-of-equilibrium phase transitions},
  \href{https://doi.org/10.1103/PhysRevD.42.2491}{\emph{Phys. Rev. D}
  {\bfseries 42} (1990) 2491}.

\bibitem{Kofman1994}
L.~Kofman, A.~Linde and A.A.~Starobinsky, \emph{Reheating after inflation},
  \href{https://doi.org/10.1103/PhysRevLett.73.3195}{\emph{Phys. Rev. Lett.}
  {\bfseries 73} (1994) 3195}.

\bibitem{Shtanov1995}
Y.~Shtanov, J.~Traschen and R.~Brandenberger, \emph{Universe reheating after
  inflation}, \href{https://doi.org/10.1103/PhysRevD.51.5438}{\emph{Phys. Rev.
  D} {\bfseries 51} (1995) 5438}.

\bibitem{Kofman1997}
L.~Kofman, A.~Linde and A.A.~Starobinsky, \emph{Towards the theory of reheating
  after inflation}, \href{https://doi.org/10.1103/PhysRevD.56.3258}{\emph{Phys.
  Rev. D} {\bfseries 56} (1997) 3258}.

\bibitem{Khlebnikov1996}
S.Y.~Khlebnikov and I.I.~Tkachev, \emph{Classical decay of the inflaton},
  \href{https://doi.org/10.1103/PhysRevLett.77.219}{\emph{Phys. Rev. Lett.}
  {\bfseries 77} (1996) 219}.

\bibitem{Bezrukov2008}
F.~Bezrukov and M.~Shaposhnikov, \emph{The standard model higgs boson as the
  inflaton},
  \href{https://doi.org/https://doi.org/10.1016/j.physletb.2007.11.072}{\emph{Physics
  Letters B} {\bfseries 659} (2008) 703 }.

\bibitem{Davies1977}
P.~Davies, S.~Fulling, S.~Christensen and T.~Bunch, \emph{Energy-momentum
  tensor of a massless scalar quantum field in a robertson-walker universe},
  \href{https://doi.org/https://doi.org/10.1016/0003-4916(77)90167-1}{\emph{Annals
  of Physics} {\bfseries 109} (1977) 108 }.

\bibitem{Bunch1977}
T.S.~Bunch and P.C.W.~Davies, \emph{Stress tensor and conformal anomalies for
  massless fields in a robertson-walker universe}, {\emph{Proceedings of the
  Royal Society of London. Series A, Mathematical and Physical Sciences}
  {\bfseries 356} (1977) 569}.

\bibitem{Birrell1982}
N.D.~Birrell and P.C.W.~Davies, \emph{Quantum Fields in Curved Space},
  Cambridge Monographs on Mathematical Physics, Cambridge University Press
  (1982),
  \href{https://doi.org/10.1017/CBO9780511622632}{10.1017/CBO9780511622632}.

\bibitem{Parker1968}
L.~Parker, \emph{Particle creation in expanding universes},
  \href{https://doi.org/10.1103/PhysRevLett.21.562}{\emph{Phys. Rev. Lett.}
  {\bfseries 21} (1968) 562}.

\bibitem{Parker1969}
L.~Parker, \emph{Quantized fields and particle creation in expanding universes.
  i}, \href{https://doi.org/10.1103/PhysRev.183.1057}{\emph{Phys. Rev.}
  {\bfseries 183} (1969) 1057}.

\bibitem{Parker1971}
L.~Parker, \emph{Quantized fields and particle creation in expanding universes.
  ii}, \href{https://doi.org/10.1103/PhysRevD.3.346}{\emph{Phys. Rev. D}
  {\bfseries 3} (1971) 346}.

\bibitem{Zeldovich1971}
Y.~Zeldovich and A.A.~Starobinsky, \emph{{Particle production and vacuum
  polarization in an anisotropic gravitational field}}, {\emph{Sov. Phys. JETP}
  {\bfseries 34} (1972) 1159}.

\bibitem{Zeldovich1977}
Y.B.~Zel'dovich and A.A.~Starobinskii, \emph{Rate of particle production in
  gravitational fields}, {\emph{JETP Lett.} {\bfseries 26} (1977) }.

\bibitem{Brout1980}
R.~Brout, F.~Englert, J.-M.~Frère, E.~Gunzig, P.~Nardone, C.~Truffin et~al.,
  \emph{Cosmogenesis and the origin of the fundamental length scale},
  \href{https://doi.org/https://doi.org/10.1016/0550-3213(80)90149-2}{\emph{Nuclear
  Physics B} {\bfseries 170} (1980) 228 }.

\bibitem{Starobinsky1984}
A.A.~Starobinsky, \emph{Nonsingular model of the universe with the
  quantum-gravitational de sitter stage and its observational consequences},
  in \emph{Quantum Gravity}, M.A.~Markov and P.C.~West, eds., (Boston, MA),
  pp.~103--128, Springer US (1984),
  \href{https://doi.org/10.1007/978-1-4613-2701-1_8}{DOI}.

\bibitem{Ford1987}
L.H.~Ford, \emph{Gravitational particle creation and inflation},
  \href{https://doi.org/10.1103/PhysRevD.35.2955}{\emph{Phys. Rev. D}
  {\bfseries 35} (1987) 2955}.

\bibitem{Mijic1986}
M.B.~Miji\ifmmode~\acute{c}\else \'{c}\fi{}, M.S.~Morris and W.-M.~Suen,
  \emph{The ${R}^{2}$ cosmology: Inflation without a phase transition},
  \href{https://doi.org/10.1103/PhysRevD.34.2934}{\emph{Phys. Rev. D}
  {\bfseries 34} (1986) 2934}.

\bibitem{Vilenkin1985}
A.~Vilenkin, \emph{Classical and quantum cosmology of the starobinsky
  inflationary model},
  \href{https://doi.org/10.1103/PhysRevD.32.2511}{\emph{Phys. Rev. D}
  {\bfseries 32} (1985) 2511}.

\bibitem{Appleby2010}
S.A.~Appleby, R.A.~Battye and A.A.~Starobinsky, \emph{Curing singularities in
  cosmological evolution {ofF}(r) gravity},
  \href{https://doi.org/10.1088/1475-7516/2010/06/005}{\emph{Journal of
  Cosmology and Astroparticle Physics} {\bfseries 2010} (2010) 005}.

\bibitem{Motohashi2012}
H.~Motohashi and A.~Nishizawa, \emph{Reheating after $f(r)$ inflation},
  \href{https://doi.org/10.1103/PhysRevD.86.083514}{\emph{Phys. Rev. D}
  {\bfseries 86} (2012) 083514}.

\bibitem{Shore1980}
G.M.~Shore, \emph{Radiatively induced spontaneous symmetry breaking and phase
  transitions in curved spacetime},
  \href{https://doi.org/https://doi.org/10.1016/0003-4916(80)90326-7}{\emph{Annals
  of Physics} {\bfseries 128} (1980) 376 }.

\bibitem{Pati1973}
J.C.~Pati and A.~Salam, \emph{Unified lepton-hadron symmetry and a gauge theory
  of the basic interactions},
  \href{https://doi.org/10.1103/PhysRevD.8.1240}{\emph{Phys. Rev. D} {\bfseries
  8} (1973) 1240}.

\bibitem{Georgi1974a}
H.~Georgi and S.L.~Glashow, \emph{Unity of all elementary-particle forces},
  \href{https://doi.org/10.1103/PhysRevLett.32.438}{\emph{Phys. Rev. Lett.}
  {\bfseries 32} (1974) 438}.

\bibitem{Georgi1974b}
H.~Georgi, H.R.~Quinn and S.~Weinberg, \emph{Hierarchy of interactions in
  unified gauge theories},
  \href{https://doi.org/10.1103/PhysRevLett.33.451}{\emph{Phys. Rev. Lett.}
  {\bfseries 33} (1974) 451}.

\bibitem{Kibble1980}
T.~Kibble, \emph{Some implications of a cosmological phase transition},
  \href{https://doi.org/https://doi.org/10.1016/0370-1573(80)90091-5}{\emph{Physics
  Reports} {\bfseries 67} (1980) 183 }.

\bibitem{Landau1982}
L.~Landau and E.~Lifschits, \emph{{The Classical Theory of Fields}},
  vol.~Volume 2 of \emph{Course of Theoretical Physics}, Pergamon Press, Oxford
  (1975).

\bibitem{Weinberg1972}
S.~Weinberg, \emph{Gravitation and Cosmology: Principles and Applications of
  the General Theory of Relativity}, John Wiley and Sons, New York (1972).

\bibitem{Starobinsky1983}
A.A.~Starobinskii, \emph{The perturbation spectrum evolving from a nonsingular
  initially de-sitter cosmology and the microwave background anisotropy},
  {\emph{Soviet Astronomy Letters} {\bfseries 9} (1983) 302}.

\bibitem{Linde_Book_1990}
A.D.~Linde, \emph{Particle Physics and Inflationary Cosmology}, Contemporary
  Concepts in Physics, CRC Press (1990).

\bibitem{Felice2010}
A.~De~Felice and S.~Tsujikawa, \emph{$f(r)$ theories},
  \href{https://doi.org/10.12942/lrr-2010-3}{\emph{Living Rev. Relativ.}
  {\bfseries 13} (2010) 3}.

\bibitem{Noh2001}
H.~Noh and J.-C.~Hwang, \emph{Inflationary spectra in generalized gravity:
  unified forms},
  \href{https://doi.org/https://doi.org/10.1016/S0370-2693(01)00875-9}{\emph{Physics
  Letters B} {\bfseries 515} (2001) 231}.

\bibitem{Oikonomou2018}
V.K.~Oikonomou, \emph{Exponential inflation with $f(r)$ gravity},
  \href{https://doi.org/10.1103/PhysRevD.97.064001}{\emph{Phys. Rev. D}
  {\bfseries 97} (2018) 064001}.

\bibitem{Akrami2020}
{Akrami, Y.} and A.~et~al. 2020, \emph{Planck 2018 results - x. constraints on
  inflation}, \href{https://doi.org/10.1051/0004-6361/201833887}{\emph{A\&A}
  {\bfseries 641} (2020) A10}.

\bibitem{D'Eramo_2017}
F.~D'Eramo, N.~Fernandez and S.~Profumo, \emph{When the universe expands too
  fast: relentless dark matter},
  \href{https://doi.org/10.1088/1475-7516/2017/05/012}{\emph{Journal of
  Cosmology and Astroparticle Physics} {\bfseries 2017} (2017) 012}.

\end{thebibliography}

\providecommand{\href}[2]{#2}\begingroup\raggedright\endgroup

\end{document}